\documentclass[fleqn,10pt]{wlscirep}
\usepackage[utf8]{inputenc}
\usepackage[T1]{fontenc}
\usepackage{lineno}
\linenumbers

\title{Exploring Urban Comfort through Novel Wearables and Environmental Surveys}

\author[1,$\dag$,*]{Patrick Chwalek}
\author[1,2,$\dag$]{Sailin Zhong}
\author[1]{Nathan Perry}
\author[5]{Tianqi Liu}
\author[3]{Clayton Miller}
\author[4]{Hamed Seiied Alavi}
\author[2]{Denis Lalanne}
\author[1]{Joseph A. Paradiso}

\affil[1]{MIT Media Lab, Cambridge, MA, USA}
\affil[2]{University of Fribourg, Fribourg, CH}
\affil[3]{National University of Singapore, SG}
\affil[4]{University of Amsterdam, Amsterdam, NL}
\affil[5]{Swiss Federal Institute of Technology, Lausanne, CH}

\affil[*]{corresponding author: Patrick Chwalek (chwalek@mit.edu)}

\affil[$\dag$]{these authors contributed equally to this work}

\begin{abstract}
This study presents a comprehensive dataset capturing indoor environmental parameters, physiological responses, and subjective perceptions across three global cities. Utilizing wearable sensors, including smart eyeglasses, and a modified Cozie app, environmental and physiological data were collected, along with pre-screening, onboarding, and recurring surveys. Peripheral cues facilitated participant engagement with micro-EMA surveys, minimizing disruption over a 5-day collection period. The dataset offers insights into urban comfort dynamics, highlighting the interplay between environmental conditions, physiological responses, and subjective perceptions. Researchers can utilize this dataset to deepen their understanding of indoor environmental quality and inform the design of healthier built environments. Access to this dataset can advance indoor environmental research and contribute to the creation of more comfortable and sustainable indoor spaces.
\end{abstract}
\begin{document}

\nolinenumbers

\flushbottom
\maketitle

\thispagestyle{empty}


\section*{Background \& Summary}

Indoor environmental health and comfort have received increased attention since the onset of the most recent pandemic. In developed countries, we typically spend over 90\% of our time indoors, yet indoor environments exhibit significant variability both between and within buildings \cite{schweizer2007indoor, Saraga2020SpecialIO}. Adding complexity, individual comfort is highly subjective, influenced by factors such as local climate, personal preferences, and individual physical and psychological attributes \cite{doi:10.1080/09613218.2015.993536, Nicol2002ADAPTIVETC}. To gain deeper insights into individual exposure and preferences, comprehensive datasets ideally should encompass diverse environmental measurements closely linked to human health and comfort, along with synchronized subjective assessments.

Measuring individuals' environmental conditions and physiological responses across different contexts poses challenges. While many commercially available sensor-rich systems are readily accessible, they often abstract away pre-processing and system intricacies from the user. Although technologies exist for monitoring physiological parameters (e.g., smartwatches), there is a scarcity of solutions optimized for continuous environmental monitoring throughout the day. The placement of environmental sensors around users is a topic of debate and depends on the specific characteristic being measured \cite{Serrano2023AdequacyOS, Yun2023OptimalSP}. For instance, for gases related to human breathing dynamics, proximity to the mouth is optimal, whereas temperature sensors afford more flexibility in placement. However, placing sensors near or on the face may present logistical constraints and discomfort for users, which could impact data accuracy. Additionally, longitudinal studies of environmental exposure should encompass periods spent outside buildings and during commuting, as these factors can vary significantly based on geographical location, local events, and transportation modes \cite{Bruin2004PersonalCM}.

Various methods exist in the literature and practice for collecting subjective data on users' comfort and perception of their environments. A recent example is the Cozie iOS and Apple Watch application, offering an open-source approach for collecting survey data in a reliable, longitudinal, and non-intrusive manner \cite{Tartarini2022CozieAA}. The application incorporates a dynamic micro-ecological momentary assessment (EMA) question flow, capturing users' location and discomforts, including thermal comfort and noise, with the flexibility to tailor assessments for specific data collection needs.

When seeking accurate sensor measurements to correlate with user perception or exposure (e.g., light levels, temperature, humidity, air quality), smart glasses emerge as an attractive option. Positioned near the eyes, ears, mouth, and nose—critical areas for visual and auditory perception, as well as inhalation and exhalation of contaminants—smart glasses offer promising avenues for data collection. For our study, we utilized the AirSpec Platform, designed for both quantitative and qualitative data collection \cite{Chwalek2023AirSpecAS}. Based on a previous smart eyeglass design, AirSpecs have been evaluated for physical comfort and appearance, demonstrating equivalence to other popular commercial smart eyeglass systems \cite{Chwalek2023CaptivatesAS}.

In this paper, we present our data collection efforts across three geographical regions using the AirSpecs platform \cite{Chwalek2023AirSpecAS} in conjunction with a custom iOS and Apple Watch application built for AirSpecs with inherited traits from the Cozie app \cite{Tartarini2022CozieAA}. We outline our data collection methodology, summarize dataset variance, provide access instructions, and offer guidance on utilization.

\section*{Methods}

\subsection*{Experimental Design}

\begin{figure}[ht]
\centering
\includegraphics[width=\linewidth]{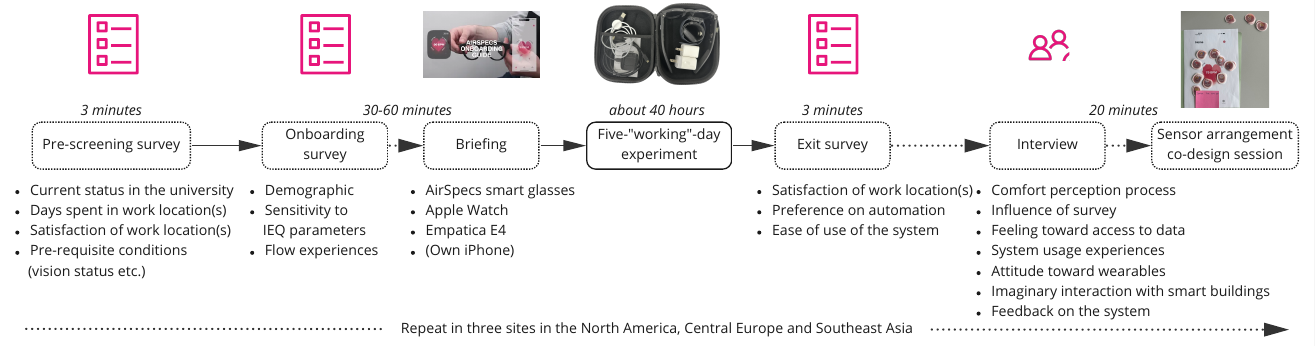}
\caption{Overview of study design.}
\label{fig:study_design}
\end{figure}

The study encompassed three geographical regions to diversify cultural backgrounds and climatic conditions: Boston, Massachusetts, USA, in March/April 2023 (Site 1); Fribourg, Switzerland, in May/June 2023 (Site 2); and Singapore in June/July 2023 (Site 3). The experiment was approved by the Institutional Review Board (IRB) of each study site, specifically, the Massachusetts Institute of Technology's IRB (2301000858), the University of Fribourg's IRB (2023-826-R2), and the National University of Singapore's IRB (NUS-IRB-2023-135). Ten participants were recruited via lab-wide email advertising at each location, and data collection occurred continuously over five days during working hours. Collection mechanisms involved pre-screening, onboarding, and recurring Ecological Momentary Assessment (EMA) surveys adapted from the Cozie iOS application \cite{Tartarini2022CozieAA}, Apple Watch, Empatica E4 wristband, AirSpecs \cite{Chwalek2023AirSpecAS}, and an exit interview (see Figure \ref{fig:study_design}).

In total, 30 participants were selected, comprising 14 women, 14 men, 1 non-binary/third gender participant, and 1 who preferred not to disclose, aged between 21 and 52 (refer to Table \ref{tab:demographics}). Participants received compensation totaling 150 local currency vouchers (equivalent to \$112-172 USD) for the entire study period, including onboarding and interview sessions. Twenty-nine participants used their own iPhones, with eight also utilizing their own Apple Watches. Others were provided with an Apple Watch 7/8 and an iPhone SE (2nd Gen). Approximately 26.6\% and 36.6\% of participants reported being slightly to extremely unsatisfied with their office and home environments, respectively. Twenty-nine participants completed five-day studies, while one completed a three-day study due to contact lens issues.

The pre-screening survey covered university status, time spent in work locations, satisfaction with work locations, and prerequisites such as vision status. The onboarding survey included demographic information and sensitivity to Indoor Environmental Quality (IEQ) parameters.

\begin{figure*}[hbpt!]
    \centering
    \includegraphics[width=0.7\textwidth]{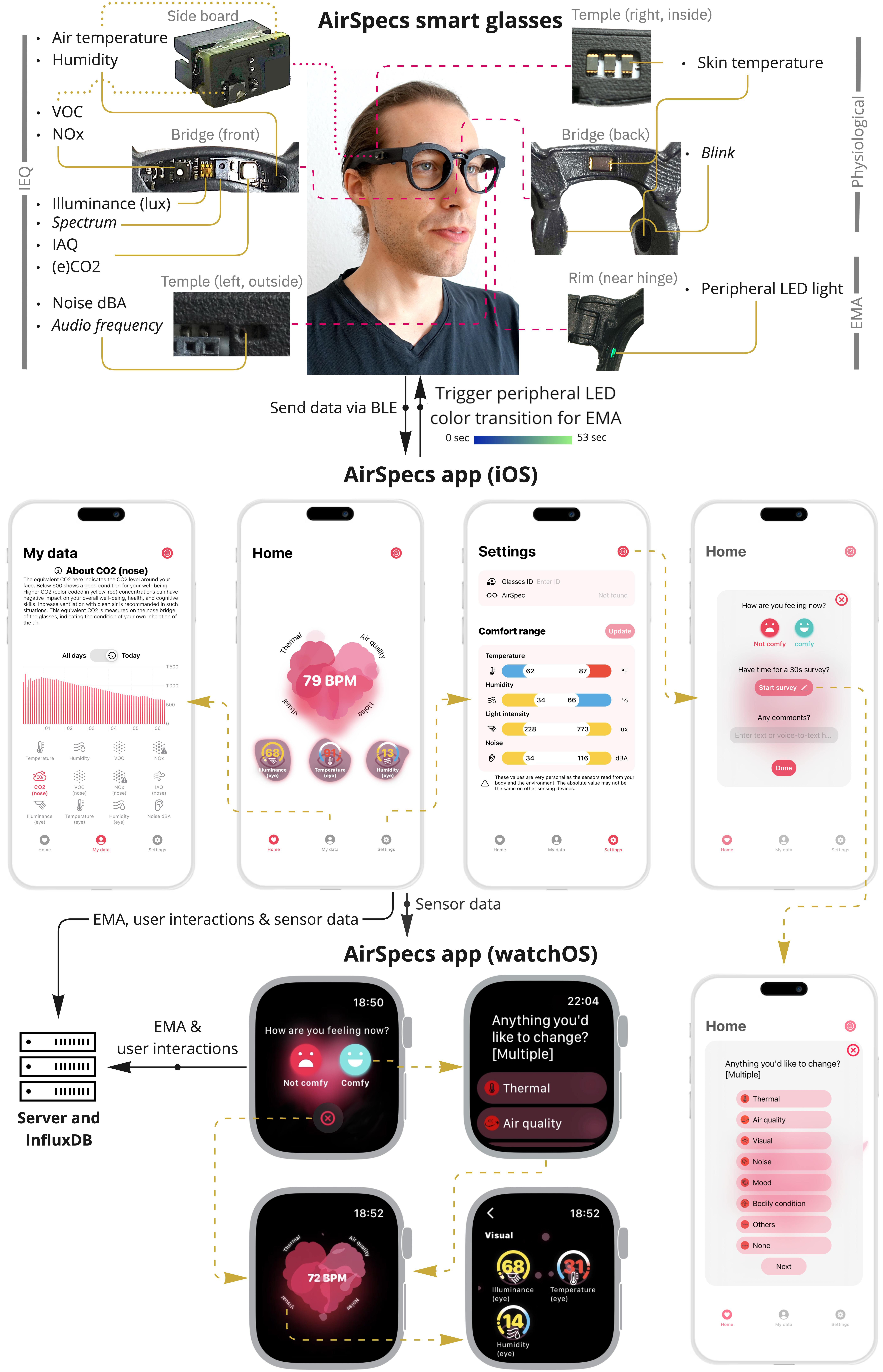}
    \caption{System overview of AirSpecs device and apps. The sensor readings of the device in the top figure (except those in \textit{italic}) are accessible by the user. Most readings are presented in raw form, except that skin temperature data was used to estimate cognitive load. Screenshots of the AirSpecs apps design in iOS and the corresponding watchOS are shown, including home, historical records, settings, and the survey. The watchOS app aims to provide spontaneous information, so we eliminated the historical records and settings screens in its design. Two sets of sensors that measure air temperature, humidity, VOC, and NOx are located in the left temple (on a sideboard) and bridge, considering that measurements near the breathing area and the surrounding environment can differ.}
    \label{fig:research_tool}
\end{figure*}

The study leveraged the AirSpecs \cite{Chwalek2023AirSpecAS} platform to collect local environmental and physiological data. The platform integrated various sensors configured at fixed sample rates and resolutions (see Table \ref{tab:sensor_location}). Data were streamed via Bluetooth Low Energy (BLE) through a custom iOS application, allowing users to view real-time data and interact with micro-EMA surveys (see Figure \ref{fig:research_tool}). All data were forwarded to an external server for monitoring by researchers to prevent data loss.

\begin{figure}[htbp]
    \centering
    \rotatebox{270}{\includegraphics[width=1.25\textwidth]{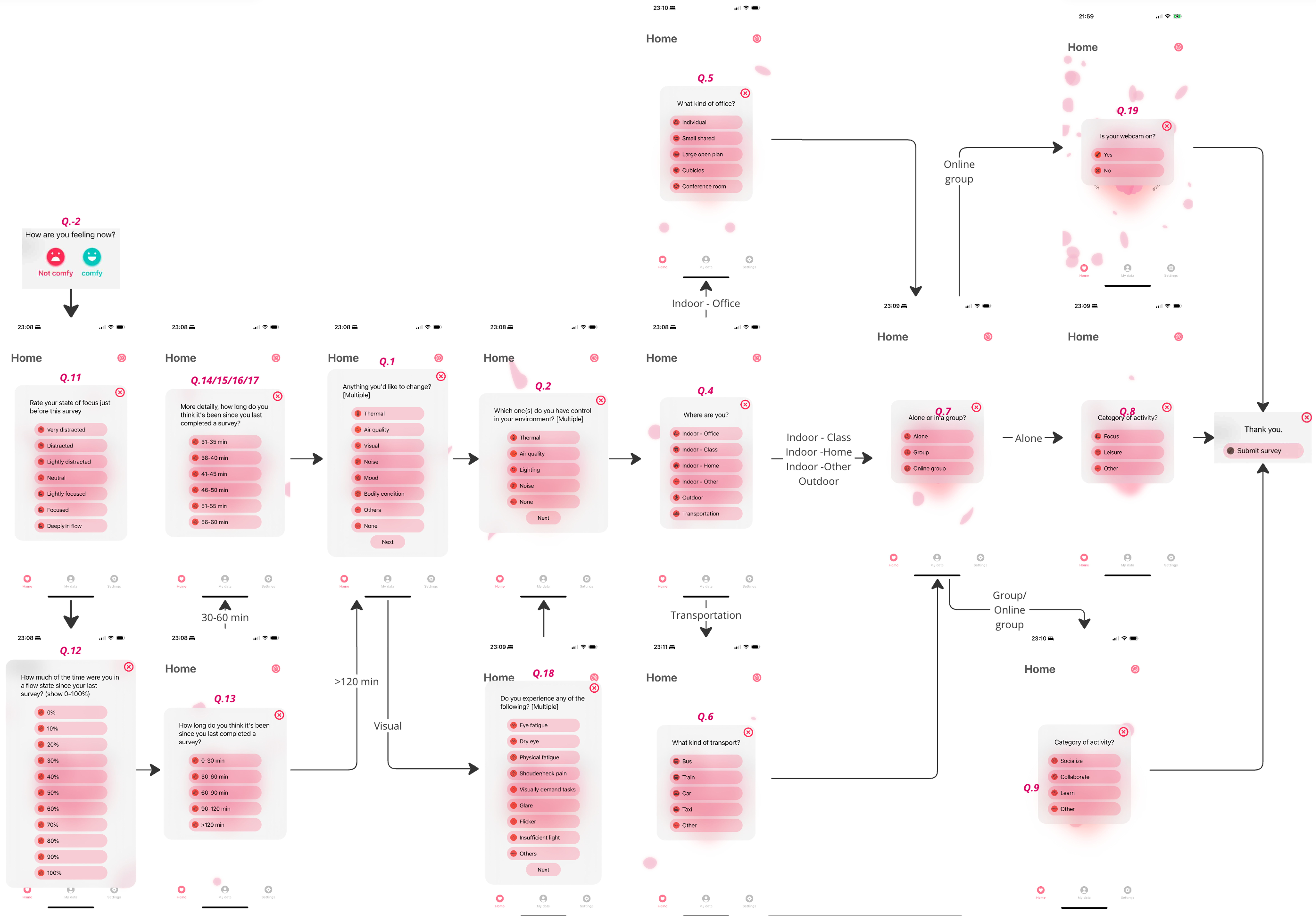}}
    \caption{The question flow is designed based on the micro-ecological momentary assessment (micro-EMA) question flows for building occupants' experience of space \cite{miller2022towards}.}
    \label{in-the-wild-survey}
\end{figure}

To prompt users to take surveys, peripheral cues were employed to reduce disruption to natural experiences. Inspired by Ramsay and Paradiso's work on using a slowly changing peripheral LED as a secondary task \cite{10.1145/3505284.3532984}, the study utilized LEDs built into AirSpecs to signal survey times at random intervals between 1 and 1.5 hours. The LED gradually transitioned from blue to green over 53 seconds to indicate survey availability (see Figure \ref{in-the-wild-survey}). If users did not respond within 15 minutes, a vibration notification was sent to their Apple Watch. Once users engaged with the micro-EMA survey, the LED reverted to blue until the next survey interval. The micro-EMA survey focused on introspective aspects, querying perceived focus level, time perception related to flow states, current context, and sources of discomfort.







 

\section*{Data Records}

The data records collected from four sources (AirSpecs, AirSpecs App, Cozie App, and Empatica E4) are synchronized using a UTC timestamp and a unique participant ID assigned to each participant. The participants' experiment schedule, along with pre-screening, demographic information, exit surveys, and open coding of sensor rearrangement co-design, are included in this dataset (\texttt{participants.csv} and \texttt{consolidated\_atlas.csv}). The meanings of the columns in the consolidated data, as well as the columns in the environmental and physiological data files, can be found in \texttt{Summary\_of\_derived\_data.xlsx}, under the "column meanings" tab. The \texttt{consolidated\_atlas.csv} includes answers to exit interview questions, which are documented in \texttt{Exit\_survey\_and\_interview\_questions.pdf} in the repository.

The details of the raw data and pre-processed consolidated data frames for each source are as follows:

\textbf{Environmental and physiological sensing data around the user's face from AirSpecs.}
The raw sensing data were exported from InfluxDB and stored in CSV format, 
and were consolidated into data frames in pickle format per sensor name and location. All of these data frames share timestamps, a unique participant ID, a phone ID (reflecting the connected glasses), and the experiment location. The rest of the columns reflect the environmental and physiological parameters recorded (e.g., ambient temperature, humidity, and skin temperature). The consolidated data is compressed within \texttt{AirSpec\_data.7z}.



\textbf{EMA from AirSpecs App.} We aimed to capture users' intuitive comfort perceptions using our Ecological Momentary Assessment (EMA) questionnaire. The questionnaire automatically progressed to the next question for single-answer queries (up to 11 questions), and participants were instructed to click "next" after completing multiple-choice questions (maximum of 3 questions). Recognizing that participants might be engaged in conversations or otherwise occupied, we allowed them to answer only the initial question on their comfort state (e.g., comfy vs. not comfy) to record reaction times accurately. Post-experimental interviews indicated that participants could complete the remaining questionnaire within 5 minutes, so we grouped these delayed responses with their initial reaction times if submitted within this timeframe. Otherwise, these responses were treated as voluntary additions.

The EMA is always initiated from the comfort state question triggered by an Apple Watch wrist-up event or iOS icon click, enabling spontaneous reaction time recording upon selecting a comfort state. User interactions with the app interfaces were logged to gauge interest in Indoor Environmental Quality (IEQ) dimensions. In our consolidated dataset, timestamps and corresponding clicks were matched with EMA, and reaction time data was recorded within a 10-minute window. 
We recommend working with the consolidated data due to the complexity of interpreting raw data without familiarity with the app's architecture and data transmission protocols.

We collected 1,175 micro-EMA surveys with associated reaction times across 30 participants across three sites: 352 from Site 1, 491 from Site 2, and 332 from Site 3. The average number of surveys per participant was $39.2 \pm 15.6$ (mean $\pm$ SD). Surveys were predominantly completed via the Apple Watch application (1,004) compared to the phone application (161). The results of the survey data across participants is stored in \texttt{survey\_reactionTime\_uiClick.csv}.

\textbf{Activity and physiological data at the user's non-dominant hand from Cozie App.} When worn on the non-dominant wrist, the Apple Watch facilitated effortless navigation using the dominant hand. Participants were instructed to wear the Apple Watch on their non-dominant hand and the Empatica E4 on their dominant hand. The Cozie App, being open-source, integrates its EMA function into the AirSpecs App described earlier, eliminating the need for participants to switch between apps during the experiment. Meanwhile, the Cozie App continued to run in the background, fetching Apple HealthKit data recorded by the Apple Watch and iPhone. Detailed parameters retrieved from HealthKit are outlined in the Cozie App's documentation available at \url{https://cozie-apple.app/docs/download_data/data_overview}. The data from the Cozie App is stored in pickle format (\texttt{cozie.pkl}) and can be deserialized using the Python pickle module.



\textbf{Physiological data at the user's dominant hand from Empatica E4.} Empatica E4 data were stored locally on the wristband for five days, and we downloaded original CSV files per sensing parameter per Empatica E4 session ID using its official app. To link session IDs with participants, we pre-processed these raw E4 files, aggregating all sensing parameters per participant per session based on their experiment schedule. These data can be further synchronized with other sources using unique participant IDs and timestamps. The data is compressed within \texttt{E4\_formatted.7z}.

\subsection*{Discontinuity}
During the experiment, we experienced a data storage failure that resulted in the loss of physiological data from AirSpecs for Site 2. However, survey data, as well as Empatica E4 and Apple Watch data, were preserved for all sites. We also do not have the blink data from Site 1 due to an issue with the initial firmware on the AirSpecs glasses that was resolved in an update prior to starting the other sites.

For the Cozie data, we don't have data for P8 and P30, likely due to the Cozie App being accidentally shut off by the participants for the duration of the experiment. The Empatica E4 dataset also doesn't include data for P23 and P26 for reasons unknown but likely hardware failures. 

\section*{Technical Validation}

The accuracy of individual environmental sensors (part numbers listed in Table \ref{tab:sensor_location}) can be found in their respective data sheets on the manufacturers' websites. We validated the non-contact skin temperature sensors by recording 30 consecutive measurements after stabilization and comparing them against the reference sensor, iButton® temperature loggers DS1922L (MAXIM Integrated, US) mounted at the same location (Table \ref{tab:temp_cal}). The iButton skin temperature sensors were calibrated at the Laboratory of Integrated Comfort Engineering (ICE), École Polytechnique Fédérale de Lausanne, using a Julabo CORIO CD water bath and a precision thermometer with an uncertainty of 0.015°C. The calibration resulted in an accuracy of ±0.2°C for the iButtons \cite{rida2023modeling}. Since we could not use both the reference sensor and the non-contact sensor simultaneously, we took a series of measurements alternating between the two in quick succession. We found that the average difference between the reference sensor and ours for the temple location was negligible, likely due to the large, relatively flat surface of the temple providing ideal conditions for the optical temperature sensors. The nose locations showed a larger temperature offset between the reference and our non-contact sensors, but the standard deviation was nearly the same as for the temple locations. This indicates that, aside from requiring a temperature offset, the performance should be similar across temperature sensors.

\section*{Usage Notes}
Within our data, there is variability in the length of sensor collections for each participant each day and in any discontinuities throughout a given day. This variability is due to the study being unsupervised by researchers, allowing participants to remove the sensors when performing tasks that might damage the units (e.g., swimming) or disrupt their activities. When using the data, discontinuities should be taken into consideration, and interpolation should be approached with caution. While some physiological and environmental parameters generally change slowly (e.g., face temperature, ambient temperature), others can vary abruptly (e.g., gases).

We have both quantitative and qualitative data that can be used to experiment with personalized model building and determine which sensing modalities and locations are the most optimal for a particular objective function. Researchers can also use the recorded response delay of the survey and results from prior work on relating response delay to focus levels to build more intelligent comfort models that consider human internal states\cite{10.1145/3505284.3532984}.

There are no access restrictions or no limitations on data use for our collected dataset. 



\section*{Code availability}

The AirSpec firmware code to reference specific sensor settings and system architectures can be found here: \url{https://github.com/pchwalek/airspec}. The iOS application with survey implementation can be found here: \url{https://github.com/sailinz/AirSpec_iOS}.


\bibliography{sample}




\section*{Author contributions statement}

P.C. and S.Z. conceived the experiment, P.C. designed and built AirSpec smart eyeglasses, S.Z. designed AirSpec iOS and Apple Watch applications, P.C. and S.Z evaluated the glasses and conducted initial pilot studies, P.C. and S.Z. conducted the experiment at Site 1, S.Z. conducted experiments at Site 2 and 3, N.P. created database for real-time and long-term storage. C.M., D.L., H.S.A., and J.A.P. advised on the project. 

\section*{Competing interests} 

The authors declare no competing interests.

\section*{Figures \& Tables}

\renewcommand{\arraystretch}{0.9} 
\begin{table*}[htbp!]
  \centering
    \begin{tabular}{llllp{0.28\linewidth}}
    \toprule
    \textbf{Parameter} & \textbf{Sensor} & \textbf{Sample rate} & \textbf{Accuracy} & \textbf{Location on glasses} \\
    \midrule
    \multicolumn{1}{p{7.835em}}{Air temperature}  & {SHT45} & {Every 5 sec} & $\pm0.1^{\circ}\text{C}$ & Temple (right, outside as side-
     \\
    Humidity &       &       & $\pm1.0\%$ & board), bridge (front) \\
    \midrule
    \multicolumn{1}{p{7.835em}}{VOC} & {SGP41} & {Every 5 sec} & $\pm15\%$ & {Temple (right, outside as side-} \\
    NOx   &       &      & $\pm50 \text{ ppb}$ & board), bridge (front)\\
    \midrule
    Iluminance (lux) & TSL27721 & Every 1 sec & - & Bridge (front) \\
    \midrule
    Spectrum & AS7341 & Every 5 sec & - & Bridge (front) \\
    \midrule
    IAQ   & {BME688} & {Every 5 sec} & $\pm15\%$ & {Bridge (front)} \\
    (e)$\mathrm{CO}_2$ &       &     & $\pm15\%$ &  \\
    \midrule
    Noise (dBA) & {ICS-43434} & \multicolumn{1}{p{10.665em}}{48000 Hz} & - & {Temple (left, outside)} \\
    Audio Frequency &       & \multicolumn{1}{p{10.665em}}{(activate 85 ms every min)} & - &  \\
    \midrule
    Skin temperature & TPIS 1S 1385 & Every 1 sec  & $\pm0.3^{\circ}\text{C}$ & Temple (right, inside), bridge (back), nose pad (right) \\
    \midrule
    Blink & QRE1113 & 1000 Hz & - & Nose pad (left) \\
    \bottomrule
    \end{tabular}%
  \caption{Summary of sensing parameters, their sampling settings, and corresponding locations on the AirSpecs device.}
  \label{tab:sensor_location}%
\end{table*}%

\begin{table}[!ht]
    \small
    \centering
    \begin{tabular}{cclllcc}
    \hline
        \textbf{PID} & \textbf{Age} & \textbf{Gender} & \textbf{Race/ethnicity} & \textbf{Occupation} & \textbf{Site} & \textbf{Work in built}\\ 
        & & & & & & \textbf{environment} \\
    \hline
        1 & 24 & Female & White & Master student & 1 & No \\ 
        2 & 25 & Female & White & University stuff & 1 & No \\ 
        3 & 29 & Non-Binary/third gender & Hispanic/Latinx & Master student & 1 & No \\ 
        4 & 24 & Male & Hispanic/Latinx, White & PhD student & 1 & No \\ 
        5 & 26 & Female & East Asian & University stuff & 1 & No \\ 
        6 & 21 & Male & White & Undergraduate student & 1 & No \\ 
        7 & 22 & Female & Asian-American & Undergraduate student & 1 & No \\ 
        8 & 32 & Male & East Asian & PhD student & 1 & Yes \\ 
        9 & 21 & Female & Hispanic/Latinx, Middle Eastern & Undergraduate student & 1 & No \\ 
        10 & 24 & Male & White & PhD student & 2 & No \\ 
        11 & 46 & Female & Hispanic/Latinx & Professor & 1 & Yes \\ 
        12 & 31 & Male & White & PhD student & 2 & No \\ 
        13 & 27 & Male & East Asian, White & PhD student & 2 & No \\ 
        14 & 27 & Male & Hispanic/Latinx, White & PhD student & 2 & No \\ 
        15 & 30 & Male & White & PhD student & 2 & No \\ 
        16 & 45 & Prefer not to say & White & Manager & 2 & No \\ 
        17 & 33 & Female & White & PhD student & 2 & No \\ 
        18 & 52 & Female & White & PhD student & 2 & No \\ 
        19 & 27 & Male & White & PhD student & 2 & Yes \\ 
        20 & 25 & Female & White & PhD student & 2 & No \\ 
        21 & 23 & Male & Southeast Asian & Undergraduate student & 3 & No \\ 
        22 & 27 & Female & East Asian & PhD student & 3 & Yes \\ 
        23 & 23 & Female & East Asian & Undergraduate student & 3 & No \\ 
        24 & 24 & Male & East Asian & PhD student & 3 & Yes \\ 
        25 & 23 & Female & East Asian & Master student & 3 & No \\ 
        26 & 35 & Male & South Asian & Master student & 3 & Yes \\ 
        27 & 26 & Female & East Asian & PhD student & 3 & No \\ 
        28 & 24 & Female & Southeast Asian & PhD student & 3 & Yes \\ 
        29 & 23 & Male & South Asian & Undergraduate student & 3 & No \\ 
        30 & 29 & Male & East Asian & Postdoc & 3 & Yes \\ \hline
    \end{tabular}
    
    \caption{30 participants were selected from 23, 14, and 45 registrations at sites 1-3 based on responses to the pre-screening survey according to three criteria: 1) prioritizing graduate students, staff, and researchers who are likely to have work tasks that require concentration, 2) no need to wear glasses or can wear contact lenses to correct their vision, and 3) not extremely satisfied with all work environments.}
    \label{tab:demographics}
\end{table}

\begin{table}[]
\centering
  \begin{tabular}{ccc}
    \hline
\textbf{Thermopile Location} & \textbf{Average (°C)} & \textbf{Standard Deviation} \\ \hline
Nose Tip                     & 0.78                  & 0.43                        \\ \hline
Nose Bridge                  & 1.05                  & 0.57                        \\ \hline
Temple (Front)               & -0.20                 & 0.40                        \\ \hline
Temple (Mid)                 & -0.10                 & 0.38                        \\ \hline
Temple (Rear)                & 0.11                  & 0.39                        \\ \hline
    \end{tabular}

 \caption{Average difference of non-contact skin temperature measurements using calibrated thermocouple as reference (n=30).}
    \label{tab:temp_cal}
    
\end{table}


\end{document}